\documentclass[preprint, pre,superscriptaddress, letterpaper, fleqn, floatfix, showpacs ]{revtex4}

\usepackage{euscript, units, amsfonts,amsmath, graphics, graphicx, dcolumn, fancyhdr}
\usepackage{times}

\usepackage[usenames]{color}

\begin{document}

\author{Lu-Jing Hou}
\affiliation{Department of Applied Mathematics, University of Waterloo, Waterloo, Ontario, Canada N2L 3G1} \affiliation{Institut
f\"{u}r Experimentelle und Angewandte Physik, Christian-Albrechts Universit\"{a}t zu Kiel, D-24098 Kiel, Germany}

\author{Z.\ L.\ Mi\v{s}kovi\'{c}}
\affiliation{Department of Applied Mathematics, University of Waterloo, Waterloo, Ontario, Canada N2L 3G1}

\title{{\Large A Gear-like Predictor-Corrector Method for Brownian Dynamics Simulation}}

\date{\today}

\begin{abstract}
We introduce a Predictor-Corrector type method suitable for performing many-particle Brownian Dynamics simulations. Since the
method goes over to the Gear's method for Molecular Dynamics simulation in the limit of vanishing friction, we refer to it as a
Gear-like algorithm. The algorithm has been tested on a one-dimensional, stochastically damped harmonic oscillator model,
showing that it can cover a wide range of friction coefficients with a high-order accuracy, excellent stability, and a very
small energy drift on the long time scales.
\end{abstract}

\received{}

\published{}

 \maketitle

\section{Introduction}
Brownian Dynamics (BD) simulation method \cite{Ermak,Allen,vanGunsteren1982,Allen1989} has been widely used in studying problems
in many dispersed systems, such as polymer solutions \cite{Ottinger}, colloidal suspensions \cite{Hutter1999,Liu2003,Chen2004},
and, more recently, complex plasmas \cite{Zheng1995,Hoffmann2000,Hou2006, LJH, Hou2008}, and has been particularly regarded as
the "mainstay" of colloid modeling over the past two decades.

BD method may be regarded as a generalization of the usual Molecular Dynamics (MD) method, which is a viewpoint that will be
held throughout the present paper. As the MD method is based on Newton's equations of motion, the BD method is based on their
stochastic generalization, namely, the Langevin equation and its integral (or Kramers equation written in Hamilton form),
\begin{eqnarray}
\frac{d}{dt}\mathbf{r}&=&\mathbf{v} \nonumber \\
\frac{d}{dt}\mathbf{v}&=&-\gamma \mathbf{v}+\frac{1}{m}\mathbf{F}+\mathbf{A}(t). \label{eqlangevin}
\end{eqnarray}
Here, as usual, $m$, $\mathbf{v}$ and $\mathbf{r}$ are respectively the mass, velocity and the position of a Brownian particle,
and $\mathbf{F}$ is the systematic (deterministic) force resulting from external sources and/or from the inter-particle
interactions in a many-particle system of Brownian particles interacting with the ambient gas of light particles. What is
different from Newton's equations is the appearance of dynamic friction, $-\gamma \mathbf{v}$, and the stochastic (Brownian)
acceleration, $\mathbf{A}(t)$. As already pointed out by Langevin, those two terms represent complementary effects of the same
microscopic phenomenon: numerous, frequent collisions between the Brownian particle and molecules in the surrounding medium.
While friction represents an average effect of those collisions, the stochastic acceleration represents fluctuations due to the
discreteness of collisions with the ambient particles, and is represented by a delta-correlated Gaussian white noise. When the
system is in thermal equilibrium, the friction and the stochastic acceleration are related to the ambient temperature via
fluctuation-dissipation theorem. Note that the friction coefficient $\gamma$ is usually regarded as constant \cite{Allen1989}.

The Langevin equation (\ref{eqlangevin}) may be integrated numerically in a manner similar to that used for Newton's equations
in MD simulations \cite{Berendsen1986,Allen1989,Leimkuhler2005,Dunweg2003, Dunweg2008}, and the resulting technique is usually
called Brownian Dynamics. However, when compared to the well-established MD techniques, algorithms for conducting BD simulations
are considerably less well developed because of their stochastic nature. It should be noted that, besides the term Brownian
Dynamics, one finds in literature other names that are used to designate various techniques of numerical integration of Langevin
equation, such as Langevin Dynamics, or Langevin Molecular Dynamics, depending on the background, area of application, or simply
preferences of various authors \cite{Dunweg2003}. It is beyond the scope of this paper to discuss all such techniques and subtle
differences among them in the derivation of the simulation formulae and/or in their implementations. Therefore, to be specific,
we are going to stick to the definitions used by Allen and Tildesley \cite{Allen1989}, and to the strategies proposed by Ermak
\textit{et al.} \emph{et al} \cite{Ermak}, Allen \cite{Allen}, van Gunsteren and Berendsen \cite{vanGunsteren1982}, which are so
far the most popular BD methods used in practice.

A straightforward method of conducting the BD simulation based on Eq.\ (\ref{eqlangevin}) was developed by Ermak \emph{et al.}
\cite{Ermak}, in which equations of motion are integrated over a time interval $t$ under the assumption that the systematic
force $\mathbf{F}$ remains approximately constant over that interval. Such method is often referred to as an Euler-like method
in the sense that it goes over to the Euler's scheme of standard MD simulation, in the limit of an infinitesimally small
$\gamma$. In spite of the low efficiency of Euler's method in the MD simulation, the Euler-like method in BD simulation seems to
be very robust for very large $\gamma$, and is still widely used in practice because of its simplicity
\cite{Ottinger,Liu2003,Chen2004}. However, very small time steps are required in this method to achieve bearable inaccuracy and
energy drift, as will be demonstrated below.

Ermak's method can be extended to higher order schemes by keeping higher-order terms in a Taylor series of the deterministic
force $\mathbf{F}$ or, equivalently in the acceleration $\mathbf{a}\equiv\mathbf{F}/m$, as follows
\begin{equation}
\mathbf{a}(t) = \mathbf{a}(0) + \dot{\mathbf{a}}(0)t + \frac{1}{2!}\ddot{\mathbf{a}}(0)t^2 +
\frac{1}{3!}\dddot{\mathbf{a}}(0)t^3 + \cdots + \frac{1}{n!}\mathbf{a}^{(n)}(0)t^n + \cdots, \label{eqtaylor}
\end{equation}
where $\mathbf{a}^{(n)}$ represents the $n$th-order time derivative of $\mathbf{a}$. For example, when integrating Eq.\
(\ref{eqlangevin}), Allen \cite{Allen,Allen1989} obtained a scheme by effectively keeping up to the second-order derivative in
Eq.\ (\ref{eqtaylor}). His method is reduced to the Beeman's method for MD simulation when $\gamma\rightarrow 0$, and is
therefore appropriately referred to as a Beeman-like method for BD simulation. We note that this method has a 3rd order accuracy
for both the velocity and position, using the definition of Ref.\ \cite{Berendsen1986} in which the order of an algorithm is
given by the highest order of the time step in simulation formulae. An another algorithm for integrating Eq.\
(\ref{eqlangevin}), which also gained popularity in practice, was developed by van Gunsteren and Berendsen \cite
{vanGunsteren1982}, by using similar assumption as Allen \cite{Allen,Allen1989}. Since their algorithm is reduced to the
position Verlet algorithm for MD simulation in the limit $\gamma\rightarrow 0$, we shall refer to it as Verlet-like method for
BD simulation. It was proven \cite{Allen,Allen1989} that the Verlet-like method is numerically equivalent to the Beeman-like
method in position, whereas the latter has higher accuracy in velocity.

Over the years, much effort has been invested in searching for better and higher-order algorithms for BD simulation
\cite{Helfand1978,Greiner1988,Mannella1989,Iniesta1990,Honeycutt1992,Hershkovitz1998, Branka,
Ricci2003,Thalmann2007,Chin1989,Chin1990,Drozdov1998,Bussi2007,Forbert2000, Chin2006}. For example, similar idea of truncating
the Taylor expansion was adopted by Helfand \cite{Helfand1978}, Iniesta and Garcia de la Torre \cite{Iniesta1990}, and Honeycutt
\cite{Honeycutt1992} to obtain second-order Runge-Kutta-like methods, and more recently by Hershkovitz \cite{Hershkovitz1998} to
obtained a fourth-order Runge-Kutta-like method. However, most of those algorithms
\cite{Helfand1978,Iniesta1990,Honeycutt1992,Hershkovitz1998, Branka} are based on direct extensions of the deterministic
Runge-Kutta algorithms to include some stochastic terms \cite{Honeycutt1992}, or on solving the deterministic part of the
Langevin equation using standard Runge-Kutta package \cite{Hershkovitz1998}. Consequently, such approaches require more than one
evaluation of the deterministic force per time step, which clearly reduces their efficiency for simulating problems in
many-particle systems \cite{Berendsen1986,Allen1989,Leimkuhler2005,Dunweg2003, Dunweg2008}.

More recently, Bra\'{n}ka and Heyes \cite{Branka} developed a series of algorithms to improve the efficiency of BD simulation
based on a finite step-size expansion for stochastic differential equations (SDEs), which may be therefore also classified as
second-order Runge-Kutta-like methods. It was found that such algorithms could significantly increase the efficiency of
simulation \cite{Branka}. However, they are mainly designed for simulating over-damped systems, such as colloidal suspensions,
as they are based on the so-called position Langevin equation \cite{Branka}, in which the variation of the velocity [i.e., the
left-hand-side of the second line in Eq.\ (\ref{eqlangevin})] is neglected \cite{Helfand1978}. Such simplification appears to be
quite adequate for simulating polymer solutions and colloidal suspensions, which are always over-damped, but may not be suitable
for simulating complex plasmas, which are typically only lightly damped \cite{Zheng1995,Hoffmann2000,Hou2006, LJH,Hou2008}.

Currently, the most advanced BD algorithms \cite{Drozdov1998,Forbert2000,Chin2006} (to the best of our knowledge) are based on
operator expansions of time propagators for Fokker-Plank or Kramers equations using Trotter decomposition technique
\cite{Chin1989,Chin1990,Drozdov1998,Forbert2000, Ricci2003,Leimkuhler2005,Dunweg2003,
Dunweg2008,Chin2006,Bussi2007,Thalmann2007}. When $\gamma\rightarrow 0$, these methods reduce to the class of so-called
symplectic integrators for Hamiltonian dynamics, which have the unique properties of conserving the phase-space volume and being
time reversible. These properties result in excellent stability properties and only very small energy drift on the long time
scales \cite{Leimkuhler2005}. However, all such methods \cite{Chin1989,Chin1990,Drozdov1998,Forbert2000,
Ricci2003,Chin2006,Bussi2007,Thalmann2007} also use the standard Runge-Kutta algorithm to solve the deterministic part of
dynamics and they, too, need multiple evaluations of deterministic forces per time step. As a consequence, they can become
computationally quite expensive when used in simulation of many-particle systems, which restricts their applicability to systems
of limited size, thus justifying further quest for more efficient algorithms of a comparable order.

On the other hand, the higher-order Predictor-Corrector (PC) methods, apparently, have not attracted much attention in BD
simulation in comparison to their popularity in MD simulation \cite{Berendsen1986,Allen1989}. A PC method for BD was proposed by
\"{O}ttinger in 1996 \cite{Ottinger}, and was subsequently adopted by H\"{u}tter \cite{Hutter1999} in a simulation of colloidal
suspensions. However, in their method the first step (predictor) was just a first-order Euler-like scheme, which obviously does
not utilize the full spectrum of advantages offered by the PC method.

Consequently, we propose in the present paper a class of general purpose PC algorithms, which cover a wide range of friction
coefficient $\gamma$ with higher-order accuracy, excellent stability, and very small energy drift on the long time scales. These
methods reduce to Gear's methods for MD simulation in the limit of $\gamma\rightarrow 0$ \cite{Berendsen1986,Allen1989}. We
shall discuss here only Gear-like algorithms that go up to the fifth-order, but extensions to higher orders should be quite
straightforward, if required. The proposed Gear-like algorithms will be tested on simulation of a simple, one-particle system,
and numerical comparisons will be made mainly with the Euler-like and Beeman-like (Verlet-like) methods because all these
methods are based on a Taylor expansion of the deterministic force, Eq.\ (\ref{eqtaylor}), and they all require only one force
evaluation per time step. As such, all these methods are specifically designed for simulations of large many-particle systems.
Further discussion of advantages of the proposed algorithms over those involving higher-order Runge-Kutta-like methods
\cite{Chin1989,Chin1990,Drozdov1998,Forbert2000,Chin2006} will be provided latter in the text.

The paper is organized as follows. In Sec.\ II, we give detailed algorithms for conducting BD simulation. We next perform in
Sec.\ III numerical tests for different methods (Euler-, Beeman- and Gear-like) by using a stochastically damped harmonic
oscillator (SDHO) model \cite{Chandrasekhar1943,Lemons2002}. Discussions of possible derivatives of the proposed BD method, as
well as comparisons with the high-order Runge-Kutta-like methods are given in Sec.\ IV, which is followed by a brief conclusion
in Sec.\ V.

\section{Algorithms for BD simulation}
\subsection{General formula}
There are various ways to derive formulae for conducting BD simulation starting from Langevin equation Eq.\ (\ref{eqlangevin}).
We shall follow here the strategy adopted in Refs.\ \cite{Ermak,Allen,vanGunsteren1982,Allen1989}, and more recently in Refs.\
\cite{Lemons1999,Lemons2002}, because it is simple and straightforward, especially for readers with some background in
simulation but lacking background in SDEs. Following the technique of integrating the Langevin equation which is elaborated in
Ref.\ \cite{Lemons2002}, one realizes that Eq.\ (\ref{eqlangevin}) can be integrated exactly over a short time, based on certain
assumptions about the deterministic acceleration $\mathbf{a}(t)$, thus giving updating formulae for BD simulation. The resulting
formulae emphasize the fact that, under the assumptions that the stochastic acceleration in Eq.\ (\ref{eqlangevin}) is a
Gaussian white noise and that $\gamma$ is constant, the two dynamic variables, $\mathbf{v}(t)$ and $\mathbf{r}(t)$, are actually
normally distributed random variables themselves [or Gaussian random variables (GRVs)]. Consequently, according to the
\emph{Normal linear transform theorem} \cite{Lemons1999,Lemons2002}, $\mathbf{v}(t)$ and $\mathbf{r}(t)$ are completely
determined by their instantaneous means (vectors) and variances (scalars), and can be expressed, respectively, as follows
\begin{eqnarray}
\mathbf{v}(t) &=& \text{mean}\{\mathbf{v}(t)\} +
\sqrt{\text{var}\{\mathbf{v}(t)\}}\,\mathbf{N_v}(0,1) \nonumber \\
\mathbf{r}(t) &=& \text{mean}\{\mathbf{r}(t)\} + \sqrt{\text{var}\{\mathbf{r}(t)\}}\,\mathbf{N_r}(0,1). \label{eqnormal}
\end{eqnarray}
Here, $\mathbf{N}(0,1)$ is a short-hand notation for a random vector [$\mathbf{N}=\{N_x,N_y,N_z\}$ in three-dimensions (3D) or
$\mathbf{N}=\{N_x,N_y\}$ in two-dimensions (2D)], whose components are mutually independent, standard (or unit) normal
distributions (i.e., having the mean $0$ and variance 1). The subscripts $\mathbf{v}$ and $\mathbf{r}$ in Eq.\ (\ref{eqnormal})
emphasize that the two random normal vectors are associated with the velocity and position, respectively, and that they are
different but correlated, as will be shown below. In the following, for the sake of simplicity, we shall occasionally denote the
means by angular brackets, $\langle \mathbf{v}\rangle$ and $\langle \mathbf{r}\rangle$, and use standard deviations
$\sigma_{v}=\sqrt{\text{var}\{\mathbf{v}\}}$ and $\sigma_{r}=\sqrt{\text{var}\{\mathbf{r}\}}$, instead of variances.

We emphasize the significance of Eq.\ (\ref{eqnormal}) because it provides the general updating formula for BD simulation.
Namely, with Eq.\ (\ref{eqnormal}), the process of solving the stochastic differential equation Eq.\ (\ref{eqlangevin}) and
obtaining the two random variables $\mathbf{v}(t)$ and $\mathbf{r}(t)$ becomes simply a matter of determining the two sets of
deterministic quantities: the means and the variances of $\mathbf{v}(t)$ and $\mathbf{r}(t)$, as well as the covariance between
$\mathbf{v}(t)$ and $\mathbf{r}(t)$. Most importantly, the variances and covariances can be shown to be independent of the form
of deterministic acceleration $\mathbf{a}(t)$. For variances, we find \cite{Ermak, Allen, Allen1989,Lemons1999,Lemons2002}:
\begin{eqnarray}
\sigma_{v}&=&\sqrt{\text{var}\{\mathbf{v}\}}=\sqrt{\frac{k_{B}T}{m}\left(1-e^{-2\gamma t}\right)} \nonumber \\
\sigma_{r}&=&\sqrt{\text{var}\{\mathbf{r}\}}=t\sqrt{\frac{2k_{B}T}{m\gamma t}\left(1-2\frac{1-e^{-\gamma t}}{\gamma
t}+\frac{1-e^{-2\gamma t }}{2\gamma t}\right)} , \label{eqvar}
\end{eqnarray}
where $k_{B}$ is the Boltzmann constant and $T$ the temperature of the surrounding medium. In passing, we note that, when a
Brownian particle is in equilibrium with the medium, $T$ also defines its kinetic temperature. However, a Brownian particle may
have a kinetic temperature different from $T$, which opens the possibility of simulating non-equilibrium processes by using BD,
such as particle stopping in a medium \cite{Hou2006}.

Further, the only non-zero covariances are $\text{cov}\{v_x,x\} = \text{cov}\{v_y,y\} = \text{cov}\{v_z,z\} \equiv
\text{cov}\{v,r\}$, and are given by (note, $\text{cov}\{v,r\} \equiv \langle vr\rangle- \langle v\rangle \langle r\rangle$ )
\cite{Lemons1999,Lemons2002}:
\begin{eqnarray}
\text{cov}\{v,r\}={t\frac{k_BT}{m\gamma t}\left (1-2e^{-\gamma t}+e^{-2\gamma t} \right )} .\label{eqcov}
\end{eqnarray}
Since the velocity and position are not independent, but rather jointly distributed normal variables, the updating formulae Eq.\
(\ref{eqnormal}) can be further written as \cite{Lemons2002}
\begin{eqnarray}
\mathbf{r}(t) & = & \langle\mathbf{r}\rangle + b_{1}\mathbf{N_{1}}(0,1)
+ b_{2}\mathbf{N_{2}}(0,1), \nonumber \\
\mathbf{v}(t) & = & \langle\mathbf{v}\rangle + \sigma_v\mathbf{N_{1}}(0,1),  \label{equpdate}
\end{eqnarray}
with $b_1=\text{cov}\{v,r \}/\sigma_{v}$ and $b_2=\sqrt{\sigma_{r}^2- b_1^2}$, where the unit normals $\mathbf{N_{1}}(0,1)$ and
$\mathbf{N_{2}}(0,1)$ are now, by design, statistically (component-wise) independent of each other, and may be generated, e.g.,
by the Box-Muller method \cite{Box1958} on a computer.

It is worthwhile pointing out that the updating formulae Eq.\ (\ref{equpdate}) can be interpreted as a two-step algorithm:
first, one calculates the means of the velocity and the position at time $t$ and, second, one adds random increments of the
velocity and position. Similar idea may also be seen in the Fokker-Planck based methods, e.g., in Ref.\ \cite{Forbert2000}. This
notion is very important in the sense that the first step is no more than evaluation of deterministic Newton's equations with
the addition of an average friction force, so that, in principle, any algorithm suitable for MD simulation (for example, Verlet,
Beeman and even multi-step PC algorithms \cite{Berendsen1986,Allen1989}) may be used here. Moreover, with the second step being
essentially independent of the first step, it can be always completed at the end of the time step as a correction to the
previous step. More importantly, in the above two steps, the inaccuracy, or error, enters only in the first step through
assumptions made regarding the deterministic acceleration $\mathbf{a}(t)$ \cite{Berendsen1986}, whereas in the second step,
random increments of the velocity and position are added with the weights obtained from Eqs.\ (\ref{eqvar},\ref{eqcov}), which
are exact and independent of the assumptions employed in the first step.

Now, the only problem that remains is to determine the means of $\mathbf{v}(t)$ and $\mathbf{r}(t)$, which are closely dependent
on the assumptions made about $\mathbf{a}(t)$, and which result in different simulation methods. We provide details for the
Gear-like methods in the following.

\subsection{Gear-like Predictor-Corrector method}
Given the form of $\mathbf{a}(t)$, defined by the manner of how is the Taylor series in Eq.\ (\ref{eqtaylor}) truncated,
$\langle \mathbf{v}\rangle$ and $\langle \mathbf{r}\rangle$ can be obtained in a number of ways by integrating Eq.\
(\ref{eqlangevin}). For simplicity, we follow here Lemons' methodology \cite{Lemons1999,Lemons2002}, in which Eq.\
(\ref{eqlangevin}) is reduced to a set of deterministic (ordinary) differential equations by taking the statistical averages of
both sides, giving
\begin{eqnarray}
\frac{d\langle\mathbf{r}\rangle}{dt} &=& \langle\mathbf{v}\rangle  \nonumber \\
\frac{d\langle\mathbf{v}\rangle}{dt} &=& -\gamma\langle\mathbf{v}\rangle + \mathbf{a}(t) , \label{eqmean}
\end{eqnarray}
where the fact that the Brownian acceleration $\mathbf{A}(t)$ is a Gaussian white noise with the mean $\mathbf{0}$ was used. It
is also important to note that the second equation in (\ref{eqmean}) may be further simplified by letting
$\langle\mathbf{v}\rangle=e^{-\gamma t}\langle\mathbf{v}\rangle^{\prime}$, so that
\begin{equation}
e^{\gamma t}\frac{d\langle\mathbf{v}\rangle^{\prime}}{dt} =
 \mathbf{a}(t) \label{eqvel} .
\end{equation}

Eq.\ (\ref{eqmean}) can be solved exactly with the help of Eq.\ (\ref{eqvel}) and, subject to the initial conditions
$\mathbf{r}_{0}$, $\mathbf{v}_{0}$, $\mathbf{a}_{0}$, $\mathbf{\dot{a}}_{0}$, $\mathbf{\ddot{a}}_{0}$, $\mathbf{\dddot{a}}_{0}$
and $\mathbf{a}_{0}^{(n)}$ at $t=0$, it yields
\begin{eqnarray}
\langle\mathbf{r}\rangle&=&\mathbf{r}_{0} + c_{1}\mathbf{v}_{0}t + c_{2}\mathbf{a}_{0}t^2 + c_{3}\mathbf{\dot{a}}_{0}t^3 +
c_{4}\mathbf{\ddot{a}}_{0}t^4 + \cdots + c_{n}\mathbf{a}_{0}^{(n-2)}t^n +
\cdots, \nonumber \\
\langle\mathbf{v}\rangle&=&c_{0}\mathbf{v}_{0}+ c_{1}\mathbf{a}_{0}t + c_{2}\mathbf{\dot{a}}_{0}t^2 + c_{3}
\mathbf{\ddot{a}}_{0}t^3 + c_{4}\mathbf{\dddot{a}}_{0}t^4 + \cdots + c_{n}\mathbf{a}_{0}^{(n-1)}t^{n} + \cdots. \label{eqglmean}
\end{eqnarray}
It is interesting to note that the coefficients in Eq.\ (\ref{eqglmean}) obey very simple recursive relations. For $n=0$, we
have $c_{0}=\exp{(-\gamma t)}$, while for $n\geq 1$, we find
\begin{eqnarray}
c_{n} &=& \frac{1}{(n-1)!\gamma
t}\left[1-(n-1)!c_{n-1}\right],\nonumber \\
\frac{d}{dt}(c_{n}t^{n}) &=& c_{n-1}t^{n-1}.  \label{eqcn}
\end{eqnarray}
As expected, these coefficients reduce to those of a standard Taylor series when $\gamma\rightarrow 0$.

To construct the Gear-like method for BD simulation, one also needs higher order derivatives of the velocity (or position)
\cite{Gear1971}. In principle, that may be attempted directly, by taking successively time derivatives on both sides of the
second equation in Eq.\ (\ref{eqmean}), but that approach turns out to be awkward. Instead, by using the equivalent equation
Eq.\ (\ref{eqvel}), one finds
\begin{eqnarray}
e^{-\gamma t}\frac{d\langle\mathbf{v}\rangle^{\prime}}{dt} &=& \mathbf{a}(t) = \mathbf{a}_{0} + \mathbf{\dot{a}}_{0}t +
\frac{1}{2!}\mathbf{\ddot{a}}_{0}t^2 + \frac{1}{3!}\mathbf{\dddot{a}}_{0}t^3 + \cdots
 \nonumber \\
\frac{d}{dt}\left[e^{-\gamma t}\frac{d\langle\mathbf{v}\rangle^{\prime}}{dt}\right] &=& \mathbf{\dot{a}}(t) =
\mathbf{\dot{a}}_{0} + \mathbf{\ddot{a}}_{0}t + \frac{1}{2!}\mathbf{\dddot{a}}_{0}t^2 + \cdots
\nonumber \\
\frac{d^2}{dt^2}\left[e^{-\gamma t}\frac{d\langle\mathbf{v}\rangle^{\prime}}{dt}\right] &=& \mathbf{\ddot{a}}(t) =
\mathbf{\ddot{a}}_{0} + \mathbf{\dddot{a}}_{0}t  + \cdots \nonumber \\
\frac{d^3}{dt^3}\left[e^{-\gamma t}\frac{d\langle\mathbf{v}\rangle^{\prime}}{dt}\right] &=& \mathbf{\dddot{a}}(t) =
\mathbf{\dddot{a}}_{0} + \cdots \label{eqtaylorlike}
\end{eqnarray}
Eq.\ (\ref{eqtaylorlike}), together with Eq.\ (\ref{eqmean}), constitutes a Taylor-like series which can be used directly to
construct the Gear-like PC method, as follows.

In a direct analogy with simulations in deterministic systems, the Gear-like PC methods for BD simulation also include three
stages, namely, predicting, force evaluating, and correcting \cite{Gear1971,Berendsen1986,Allen1989}. The difference is that,
one has to add random displacements of the velocity and position by using Eq.\ (\ref{equpdate}) at the end of a time step to
complete the BD simulation. The basic procedure goes as follows.

\emph{Predicting:} In this stage, one has:
\begin{eqnarray}
\mathbf{r}^{P} &=&\mathbf{r}_{0} + c_{1}\mathbf{v}_{0}t + c_{2}\mathbf{a}_{0}t^2 + c_{3}\mathbf{\dot{a}}_{0}t^3 +
c_{4}\mathbf{\ddot{a}}_{0}t^4 + c_{5}\mathbf{\dddot{a}}_{0}t^5
\nonumber \\
\mathbf{v}^{P} &=&c_{0}\mathbf{v}_{0}+ c_{1}\mathbf{a}_{0}t + c_{2}\mathbf{\dot{a}}_{0}t^2 + c_{3}
\mathbf{\ddot{a}}_{0}t^3 + c_{4}\mathbf{\dddot{a}}_{0}t^4 \nonumber \\
\mathbf{a}^{P} &=& \mathbf{a}_{0} + \dot{\mathbf{a}}_{0}t + \frac{1}{2!}\ddot{\mathbf{a}}_{0}t^2 +
\frac{1}{3!}\dddot{\mathbf{a}}_{0}t^3
 \nonumber \\
 \mathbf{\dot{a}}^{P} &=&
\mathbf{\dot{a}}_{0} +
\mathbf{\ddot{a}}_{0}t + \frac{1}{2!}\mathbf{\dddot{a}}_{0}t^2 \nonumber \\
\ddot{\mathbf{a}}^{P} &=&
\mathbf{\ddot{a}}_{0} + \mathbf{\dddot{a}}_{0}t \nonumber \\
\dddot{\mathbf{a}}^{P} &=& \mathbf{\dddot{a}}_{0}, \label{eqpredictor}
\end{eqnarray}
where the superscript $P$ is to indicate that these are quantities in the predicting stage. (For simplicity, we have dropped
derivatives of $\mathbf{a}(t)$ higher than the third order, but extensions to higher orders are quite straightforward.) One may
notice in Eq.\ (\ref{eqpredictor}) that we have used in Eq.\ (\ref{eqglmean}) the means of the position and velocity instead of
using Taylor series of the position and velocity, as is normally done in the Gear method for MD simulation
\cite{Berendsen1986,Allen1989}. Other than that, the remaining parts of this stage (derivatives of the force) are essentially
the same as those in the MD simulation.

\emph{Force evaluating:} In this stage, the predicted position $\mathbf{r}^{P}(t)$ is used to obtain a new force, that is, new
acceleration $\mathbf{a}(t)$, and a difference between the predicted acceleration $\mathbf{a}^{P}(t)$ and the new acceleration
$\mathbf{a}(t)$ is formed:
\begin{equation}
\Delta\mathbf{a}=\left [\mathbf{a}(t)- \mathbf{a}^{P}(t) \right ]. \label{eqfe}
\end{equation}
It can be seen that this step is exactly the same as in MD simulation \cite{Berendsen1986,Allen1989}.

\emph{Correcting:} In this stage, the above difference term is used to correct all predicted positions and their "derivatives",
as follows:
\begin{eqnarray}
\mathbf{r}^{C} &=& \mathbf{r}^{P} + 2c_{2}\alpha_{0}\Delta \mathbf{R2} \nonumber
\\
\mathbf{v}^{C}t &=& \mathbf{v}^{P}t + c_{1}\alpha_{1}\Delta \mathbf{R2} \nonumber
\\
\frac{\mathbf{a}^{C}t^2}{2!} &=& \frac{\mathbf{a}^{P}t^2}{2!} + \alpha_{2}\Delta \mathbf{R2} \nonumber
\\
\frac{\mathbf{\dot{a}}^{C}t^3}{3!} &=& \frac{\mathbf{\dot{a}}^{P} t^3}{3!} + \alpha_{3}\Delta \mathbf{R2} \nonumber
\\
\frac{\mathbf{\ddot{a}}^{C}t^4}{4!} &=& \frac{\mathbf{\ddot{a}}^{P}t^4}{4!} + \alpha_{4}\Delta \mathbf{R2} \nonumber
\\
\frac{\mathbf{\dddot{a}}^{C}t^5}{5!} &=& \frac{\mathbf{\dddot{a}}^{P}t^5}{5!} + \alpha_{5}\Delta \mathbf{R2} \label{eqcorrector}
\end{eqnarray}
with
\begin{equation}
\Delta \mathbf{R2} \equiv \frac{\Delta\mathbf{a}t^2}{2!}, \label{eqdelta}
\end{equation}
and the coefficients given in the following table:
\begin{center}
\begin{tabular*}{0.75\textwidth}{@{\extracolsep{\fill}}|cccc|}\hline
$\alpha_{i}$ &      3th-order      &      4th-order           &     5th-order \\
\hline\hline
$\alpha_{0}$ & $\frac{1}{6}$ & $\frac{19}{120}$   & $\frac{3}{16}$ \\
\hline
$\alpha_{1}$ & $\frac{5}{6}$ & $\frac{3}{4}$      & $\frac{251}{360}$ \\
\hline
$\alpha_{2}$ &      1        &         1          &      1           \\
\hline
$\alpha_{3}$ & $\frac{1}{3}$ & $\frac{1}{2}$      & $\frac{11}{18}$ \\
\hline
$\alpha_{4}$ &       0       & $\frac{1}{12}$     & $\frac{1}{6}$ \\
\hline
$\alpha_{5}$ &       0       &         0          & $\frac{1}{60}$ \\
\hline
\end{tabular*}
\end{center}
This table is simply a reproduction of those in Refs. \cite{Gear1971,Berendsen1986,Allen1989}, and is provided here for
completeness. By using parameters in different columns, one may realize 3rd-, 4th-, and 5th-order (or 4-, 5- and 6-value)
\cite{Berendsen1986,Allen1989} Gear-like algorithms for BD simulation. Note that the first two lines in Eq.\ (\ref{eqcorrector})
are slightly different from those in MD simulation, in order to restore the damping effect on deterministic acceleration, and to
keep the consistency with the corresponding terms in Eq.\ (\ref{eqglmean}) [or the first two lines in Eq.\ (\ref{eqpredictor})
].

\emph{Adding random displacements:} The above three stages are the same as those of Gear's algorithm for MD simulation. However,
to complete the BD simulation, we have to use the updating formulae Eq.\ (\ref{equpdate}) to add random displacements of the
velocity and position. It should be noted that now the corrected values $\mathbf{r}^{C}$ and $\mathbf{v}^{C}$ should be used in
the places of $\langle\mathbf{r}\rangle$ and $\langle\mathbf{v}\rangle$ in Eq.\ (\ref{equpdate}), respectively. Also note that,
to implement the simulation, one needs the initial conditions, $\mathbf{r}_{0}$, $\mathbf{v}_{0}$, $\mathbf{a}_{0}$,
$\mathbf{\dot{a}}_{0}$, $\mathbf{\ddot{a}}_{0}$ and $\mathbf{\dddot{a}}_{0}$, at $t=0$. This poses a problem for the very first
several steps of simulation, because $\mathbf{\dot{a}}_{0}$, $\mathbf{\ddot{a}}_{0}$ and $\mathbf{\dddot{a}}_{0}$ are undefined.
The simplest way to get around this issue is to set all of them to be zero at the very first step, whereas their values will be
updated during subsequent iterations. A better way would be to start the simulation by using a Runge-Kutta procedure for the
first few steps \cite{Berendsen1986}. However, neither of these solutions of the problem will have any effect on the results in
real, many-particle simulations.

The above are the basic procedures for the Gear-like PC method for BD simulation. It should be restated that these formulae and
the simulation stages are quite similar to those in the Gear method for MD simulation of Newton's equations
\cite{Berendsen1986,Allen1989}, except for the expressions for the velocity and position in Eq.\ (\ref{eqpredictor}) and Eq.\
(\ref{eqcorrector}), and for the addition of random displacement at the end of every time step. When $\gamma\rightarrow 0$, the
Gear-like method goes over to the Gear method for MD simulation. In the following we shall show that the performance of the
proposed Gear-like method also shares many features with its counterpart in MD simulations.

\section{Testing the algorithm}
\begin{figure}
\centering
\includegraphics[trim=0 10mm 0 10mm, width=0.7\textwidth]{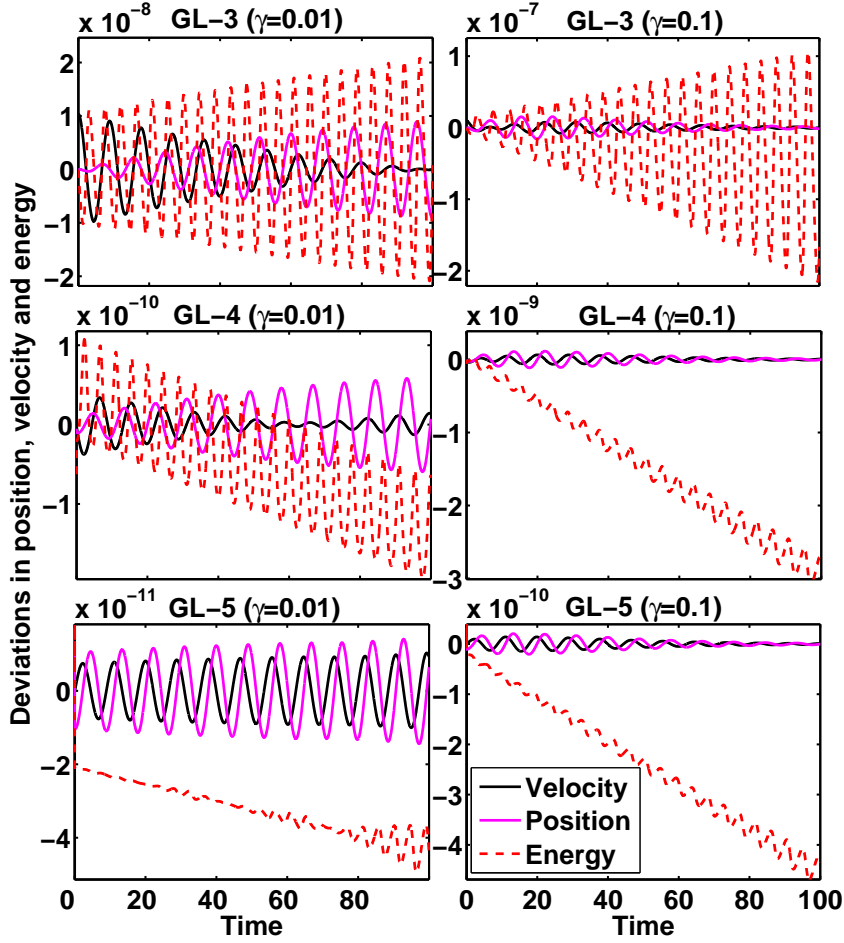}
\caption{(Color online) Deviations in the position, velocity and energy from their respective exact solutions within the first
100 time units. The left column is for $\gamma=0.01$ and the right one for $\gamma=0.1$. Here, GL-3, GL-4 and GL-5 denote the
3rd-, 4th-, and 5th-order Gear-like methods, respectively. The time step is fixed at $0.01$. The (red) dashed lines, (blue)
solid lines, and (black) solid lines show the deviations of energy, velocity and position, respectively. Note that the energy
deviation is defined relative to the exact values by $E(t)/E_{e}(t)-1$.} \label{fig_GL_error}
\end{figure}

In this section, we present some simple computational examples as testing cases for our Gear-like algorithm, and for the
previously designed Euler-like and Beeman-like (Verlet-like) algorithms. For simplicity, we shall occasionally denote the
Euler-like, Beeman-like, Verlet-like and Gear-like methods by EL, BL, VL, and GL, respectively.

We employ the model of a one-dimensional (1D) SDHO \cite{Chandrasekhar1943, Lemons2002}, in which Eq.\ (\ref{eqlangevin}) may be
simplified to
\begin{eqnarray}
\frac{d}{dt}v&=&-\gamma v - \omega_{0}^{2}r+A(t) ,
\nonumber \\
\frac{d}{dt}r&=&v  \label{eqsdho0}.
\end{eqnarray}
This model has been studied extensively and its solutions are well known \cite{Chandrasekhar1943,Lemons2002}. We have listed
Chandrasekhar's results in the Appendix A to facilitate subsequent discussion.

According to our discussion above, especially following Eq.\ (\ref{equpdate}), the task of a BD simulation is simply to predict
the position $\mathbf{r}(t)$ and velocity $\mathbf{v}(t)$ of a Brownian particle at time $t$, given the set of initial
conditions at time $0$. As previously pointed out, $\mathbf{r}(t)$ and $\mathbf{v}(t)$ are normally distributed random vectors,
whose variances and covariances are exact (at the level of Langevin equation), whereas their means are calculated numerically
according to certain approximation schemes, such as EL, BL, VL or GL methods. From this point of view, the errors come only from
evaluation of the means. In other words, the accuracy in calculating the means will decide the main performance of a BD
simulation, so we begin our tests by considering the means.

As usual \cite{Berendsen1986,Allen1989}, we firstly carry out simulations over certain periods and record, as a measure of
accuracy, the deviations of $r$ and $v$ from their exact solutions without the inclusion of random displacements. We shall also
judge the behavior of simulations by monitoring the quantity
\begin{equation}
E(t)=v^{2}(t) + \omega_0^2 r^{2}(t), \label{eqenergy}
\end{equation}
which is proportional to the total energy of the Brownian particle. Note that, because of the existence of damping, this energy
is no longer a conserved quantity, but rather decreases exponentially with the factor $\exp(-2\gamma t)$. This may be seen from
the exact solutions for the mean position and velocity listed in the Appendix C. In order to judge the energy conservation
performance of numerical methods, we normalize $E(t)$ by its exact counterpart $E_{e}(t)$, which is simply obtained from Eq.\
(\ref{eqenergy}) by substituting the exact values for the position and velocity. The resultant ratio $E(t)/E_{e}(t)$ should be a
conserved quantity with the expected value of unity.

\begin{figure}
\centering
\includegraphics[trim=0 10mm 0 10mm, width=0.5\textwidth]{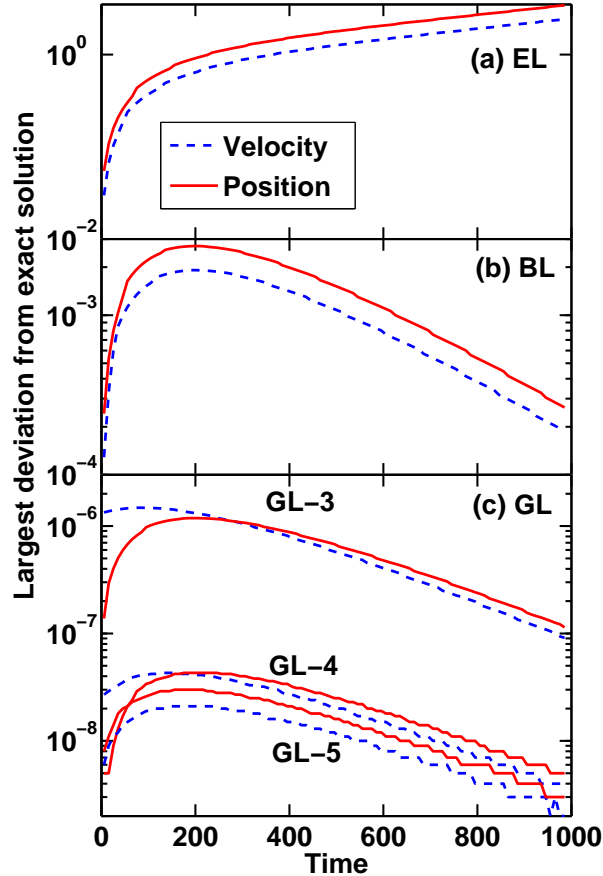}
\caption{(Color online) Amplitudes of time-dependent deviations (or the largest deviations) in the position and velocity within
1000 time units for different methods. Here, EL and BL denote the Euler-like and Beeman-like methods, respectively.
$\gamma=0.01$ and the time step of $0.05$ are used in these tests.} \label{fig_growthrate}
\end{figure}

Without loss of generality, in all ensuing simulations we set $k_{B}T=1$, $m=1$ and $\omega_{0}=\sqrt{2}/2$, and choose the
initial conditions to be $r_{0}=1$ and $v_{0}=0$. These parameters are also used in the exact results based on Appendix A.

The results of simulations for deviations in the position, velocity and energy from their exact solutions) are shown in Fig.\
\ref{fig_GL_error} for the 3rd-, 4th-, and 5th-order Gear-like methods (denoted as GL-3, GL-4 and GL-5, respectively) within the
first 100 time units. Note that the period of the damped oscillator is approximately $8.9$ for both $\gamma=0.01$ and
$\gamma=0.1$, and that the highest amplitude of velocity is approximately $\sqrt{2}/2$. The oscillatory patterns of the
deviations in position and velocity resemble approximately those of the exact solutions, but with much smaller amplitudes. The
patterns of the energy deviations have a higher frequency (doubled) and seldom oscillate around 0; rather, their oscillating
centers drift in an approximately linear manner (in a semi-logarithmic plot). The slope of this line is usually defined as the
\emph{energy drift}, which is the main criterion testing the energy conservation in an algorithm \cite{Berendsen1986,
Allen1989}. Of course, one would expect the slope to be as small as possible. Another criterion for the energy conservation is
the amplitude of oscillations in the energy deviation, which is called the \emph{noise} of the energy drift. Needless to say,
the \emph{noise} should be small, too, in accurate simulations. It is interesting to notice that some of the slopes in Fig.\
\ref{fig_GL_error} are negative, indicating that the energy is damped. This is quite different from the usual situation in the
MD simulation where damping is absent by definition, and where the slopes of \emph{energy drift} are always found positive.
This, and other aspects of \emph{energy drift} will be further discussed below.

The amplitudes of the time-dependent deviations in position and velocity are shown in Fig.\ \ref{fig_growthrate} for an extended
time period of $1000$ time units. One can easily see large differences in the amplitudes among different methods, with the
patterns of the Euler-like and Beeman-like (Verlet-like) methods looking similar to those of the Gear-like method, but
exhibiting orders of magnitude larger amplitudes than the Gear-like method. This is to be expected as, generally, all Gear-like
methods have much better accuracy than Euler-like and Beeman-like (Verlet-like) methods. It should be furthermore pointed out
that, except for deviations of the Euler-like method, which grow monotonously, all other deviations pass through broad maxima
around the time of 200 time units. This indicates that the Euler-like method is not stable, while the others are. This
illustrates yet another aspect in which BD algorithms differ from those of MD simulation, where the deviations invariably grow
with time \cite{Allen1989}. Namely, it is obvious that the finite damping $\gamma$ actually stabilizes numerical algorithms
(except for the Euler-like method). This is not surprising at all, because previous studies
\cite{Izaguirre2001,Dunweg2003,Dunweg2008} have demonstrated a possibility of stabilizing the MD simulation by introducing a
small, but finite damping. A similar idea is also well-known in the computational fluid dynamics where, for example, a small
artificial viscosity is often introduced in fluid simulations to stabilize the computation \cite{Anderson1995}.

\begin{figure}
\centering
\includegraphics[trim=0 10mm 0 10mm, width=0.7\textwidth]{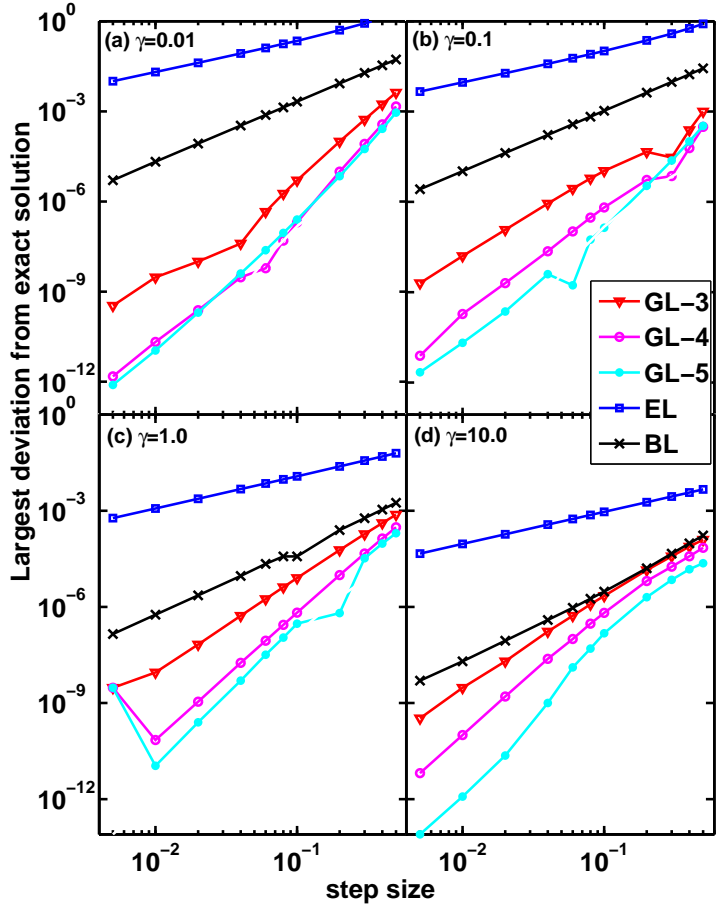}
\caption{(Color online) The largest deviation in the position from exact solution vs.\ the time step size in the first 20 time
units, for different methods and different $\gamma$s. } \label{fig_devi}
\end{figure}

\begin{figure}
\centering
\includegraphics[trim=0 10mm 0 10mm, width=0.7\textwidth]{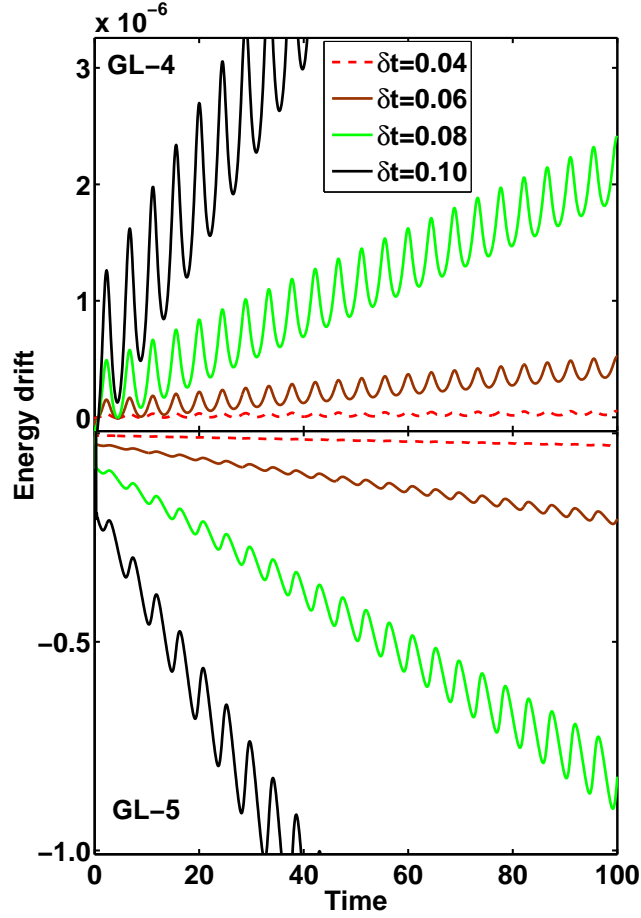}
\caption{(Color online) Time-dependent energy drift in GL-4 and GL-5 for different time step sizes $\delta t$, with
$\gamma=0.01$ being fixed.} \label{fig_GL_drift}
\end{figure}

Next, we make some comparisons involving different time steps, as that is always the key issue in both the MD and BD
simulations. Fig.\ \ref{fig_devi} shows the largest deviations (defined by the largest deviation in the first 20 units) in
position versus the size $\delta t$ of the time step for different methods and different friction coefficients $\gamma$. This
figure is plotted in a double logarithmic scale, and the curves are nearly straight lines. The slope of these lines is called
the \emph{apparent order} \cite{Berendsen1986}, illustrating the dependence of error on the time step size. Namely, if the error
is found to be proportional to $\delta t^{p}$, then the exponent $p$ is the \emph{apparent order}. It is found that for very
small $\gamma$, for example $\gamma=0.01$, as shown in Fig.\ \ref{fig_devi}(a), the \emph{apparent order}s are $p_{EL}\approx
1$, $p_{BL}\approx 2$, $p_{GL-3}\approx 3.5$, $p_{GL-4}\approx 4.2$ and $p_{GL-5}\approx 4.6$, respectively for the Euler-like,
Beeman-like, 3rd-, 4th-, and 5th-order Gear-like methods. These values are very close to those from the MD simulations where
damping is absent \cite{Berendsen1986}. When $\gamma$ increases, the absolute value of the error for Euler-like and Beeman-like
methods decreases, while $p_{EL}$ and $p_{BL}$ remain almost unchanged. On the other hand, $p_{GL-3}$, $p_{GL-4}$ and $p_{GL-5}$
slightly decrease with increasing $\gamma$. For example, for $\gamma=1$, as shown in Fig.\ \ref{fig_devi}(c), Gear-like methods
have the worst performance in accuracy: the \emph{apparent order}s become now $p_{GL-3}\approx 3.1$, $p_{GL-4}\approx 3.9$ and
$p_{GL-5}\approx 4.0$), but the amplitudes of their errors are still much smaller than those of the Beeman-like and Euler-like
methods.

We further provide a more detailed analysis of the energy conservation performances. We have already discussed the \emph{energy
drift} and \emph{noise} for Gear-like methods in Fig.\ \ref{fig_GL_error}. In particular, negative energy drifts were noticed
there for the 4th- and 5th-order Gear-like methods. However, this is not the case for all time step sizes, as illustrated in
Fig.\ \ref{fig_GL_drift}, showing the energy drifts in the 4th- and 5th-order methods for different time step-sizes. We see
that, when the step-size increases, the energy drift for GL-4 becomes increasingly positive, while that of GL-5 grows
increasingly negative, until a very large time step is used (e.g., $\delta t=0.6$, not shown in the figure). On the other hand,
the noise in the energy drift is always seen to increase with $\delta t$.

However, for GL-3, the trends in the energy drift are seen in Fig.\ \ref{fig_GL_error} to be quite different, especially for
smaller $\gamma$. Namely, the oscillations in the energy deviation seem rather symmetrical about zero, with the amplitude of the
noise being at least two orders of magnitude larger than the drift, as can be seen in the top left panel of Fig.\
\ref{fig_GL_error}. It is found that the Beeman-like method exhibits similar behavior as GL-3, which makes it very hard to do
the least-squares fitting and linear regression analysis \cite{Berendsen1986}. On the other hand, for large $\gamma$, e.g.
$\gamma>1$, the whole energy of the system is practically damped to zero in a few time units, so that it does not make much
sense to discuss the energy drift in such cases. Therefore, in the following  we shall omit the energy drift analysis for the BL
and GL-3 methods altogether, as well as for other methods in the cases of large $\gamma$.

\begin{figure}
\centering
\includegraphics[trim=0 10mm 0 10mm, width=0.7\textwidth]{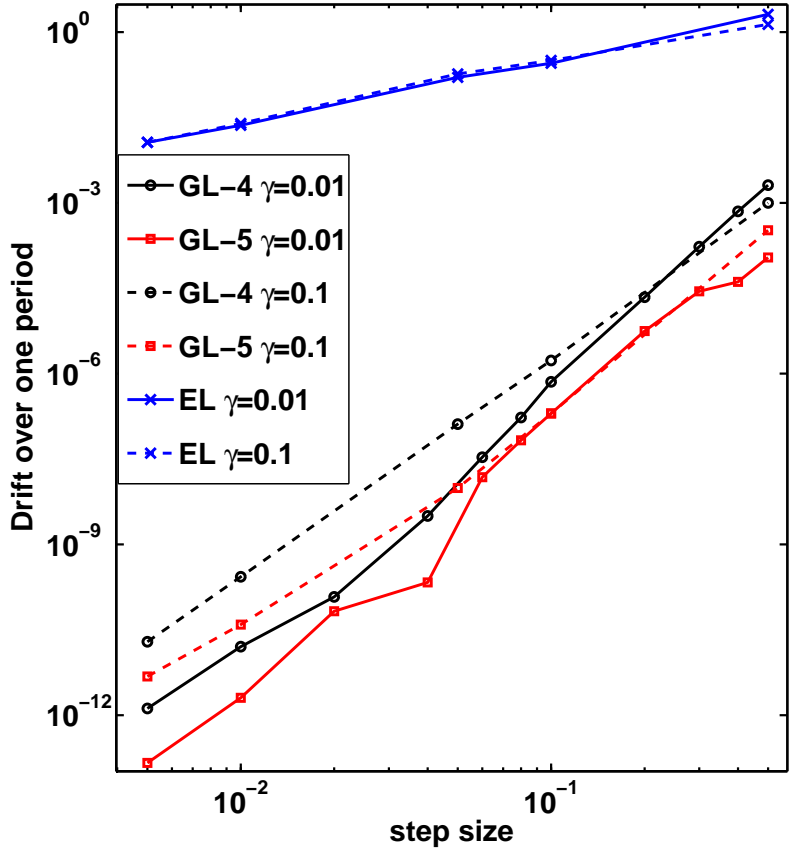}
\caption{(Color online) Energy drift in one period of oscillation vs.\ the time step size for different methods and for
$\gamma=0.01$ and $0.1$. } \label{fig_drift}
\end{figure}

\begin{figure}
\centering
\includegraphics[trim=0 10mm 0 10mm, width=0.7\textwidth]{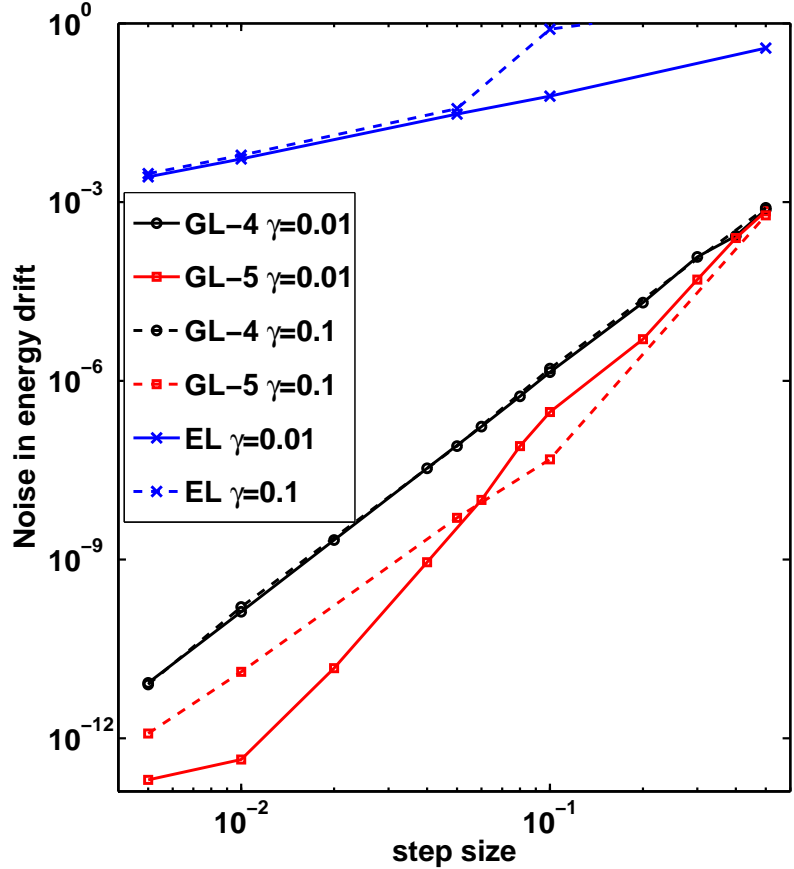}
\caption{(Color online) Noise in the energy drift (defined by the average amplitude of oscillations in the first 100 time units)
vs.\ the time step size, for different methods and for $\gamma=0.01$ and $0.1$. } \label{fig_noise}
\end{figure}

Figure \ref{fig_drift} shows the \emph{energy drift} over one period of oscillations versus the time step size
\cite{Berendsen1986} for the EL, GL-4 and GL-5 methods with two different friction coefficients $\gamma$. Similarly, Fig.\
\ref{fig_noise} shows the noise in the energy deviation (defined by the averaged amplitudes of oscillations in the first 100
time units) versus the time step size, for these same methods with the same friction coefficients $\gamma$. Since those figures
are also plotted in a double logarithmic scale, we see that both the \emph{energy drift} and the \emph{noise} increase almost
linearly with the time step size, and that the slopes of these curves are very close to the corresponding \emph{apparent
order}s, shown in Fig.\ \ref{fig_devi}. Moreover, the energy drifts for $\gamma=0.1$ in the GL-4 and GL-5 methods seem worse off
than the corresponding drifts for $\gamma=0.01$, but the magnitudes of the \emph{energy drift} and \emph{noise} are still very
small.

\begin{figure}
\centering
\includegraphics[trim=0 10mm 0 10mm, width=0.7\textwidth]{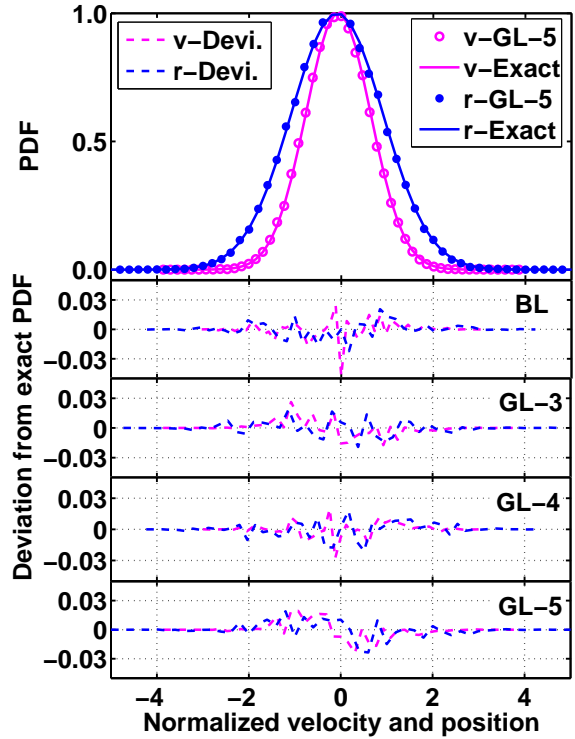}
\caption{(Color online) Solid lines show the exact probability distribution functions (PDFs) for velocity and position given by
Eq.\ (\ref{eqsdhon}), while symbols (circles and dots) show the statistics of $80,000$ samples taken from $800,000$ time units
of simulation using the G-5 method. (Note that the PDFs have been scaled up, and their peaks are now unity.) Lower panels show
the deviations of PDFs obtained in simulations from the exact ones, for the BL, GL-3, GL-4, and GL-5 methods. $\gamma=0.01$ and
$\delta t=0.1$ are used in these simulations.} \label{fig_var}
\end{figure}

In the above, we have discussed the short-time performances of various BD methods in terms of accuracy and energy drift, without
considering random displacements in the position and velocity. In the following, we perform several long-time tests including
the random terms, in order to test the statistical properties of the simulation results.

In the long time limit, the oscillator approaches thermal equilibrium with its environment, so that both its velocity and
position should become normally distributed. This may be easily seen from their exact solutions, Eq.\ (\ref{eqsdho}), or more
clearly from their long time limits, Eq.\ (\ref{eqsdhon}), which are plotted in Fig.\ \ref{fig_var}, along with the statistical
ensembles obtained from the simulation data. In this figure, solid lines show the probability distribution functions (PDFs) of
the velocity and position given by Eq.\ (\ref{eqsdhon}), whereas symbols (circles and dots) show ensembles of $80,000$ sample
points taken from $800,000$ time units of simulation. (Note that the first $2000$ time units are used to allow the oscillator
\begin{figure} \centering
\includegraphics[trim=10 50mm 10 10mm, width=0.8\textwidth]{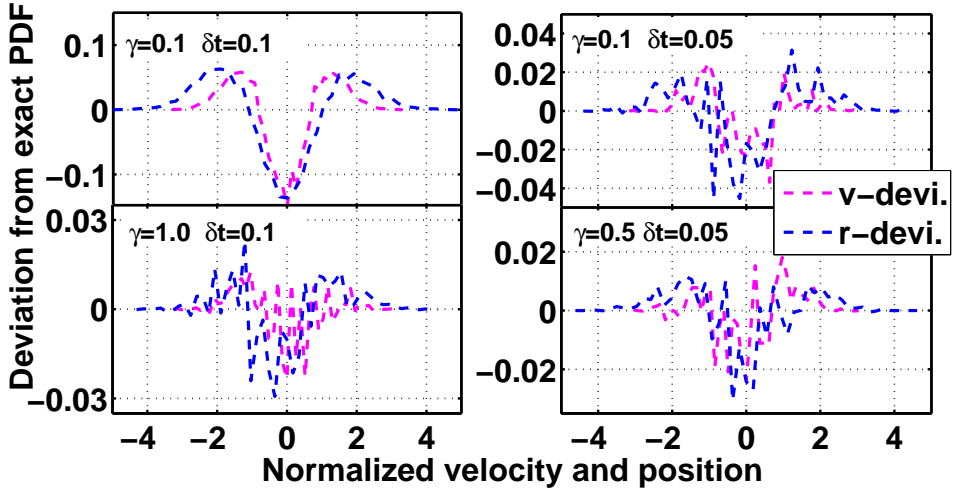}
\caption{(Color online) Deviations of PDFs for velocity and position for the EL method, with different time step sizes and
damping coefficients. Note that the PDFs have been scaled up, and their peaks are now unity, as is shown in Fig. \ref{fig_var}.}
\label{fig_el_var}
\end{figure}
to reach an equilibrium, so that data from this period are not included in the ensembles shown in Fig.\ \ref{fig_var}). The
dashed lines display deviations between the simulation data and the exact PDFs. We used $\gamma=0.01$ and $\delta t=0.1$ in
simulations, allowing us to compare the results from the BL, GL-3, GL-4 and GL-5 methods, whereas the EL method is ruled out
because it is not stable under these conditions. Since there are essentially no visible differences in the distribution
functions generated by these four methods, we only show the PDF from the GL-5 method as an example, but display all deviations.
We see that the deviations of all GL methods lie approximately within the range of $(-3\%, 3\%)$, which is the statistical error
for $80,000$ samples according to our experience. [We use the Box-Muller method to take $80,000$ samples, with the PDF given by
Eq.\ (\ref{eqsdhon}), and compare the resultant PDF with the exact PDF. This gives a deviation in the range of $(-3\%, 3\%)$.]
The deviation from the BL method, as shown in the figure, is a bit larger: within the range of $(-4\%, 4\%)$. With a smaller
time step $\delta t$ and/or larger $\gamma$, all the above methods (EL, GL-3, GL-4, and GL-5) exhibit basically the same
performance in the long time limit, and all their deviations are well within the range of the statistical error.

Generally, a larger $\gamma$ would allow a larger time step $\delta t$ to be used for the same accuracy in the long time limit,
and this is true for all methods including the EL method. For example, this method requires typically $\gamma>1$, along with the
step size of about $\delta t=0.1$, enabling it to produce deviations in the PDF comparable to the other methods, as is shown in
Fig.\ \ref{fig_el_var}.

\section{Discussion}
\subsection{Possible simplification}
\begin{figure}
\centering
\includegraphics[trim=0 10mm 0 0mm, width=0.7\textwidth]{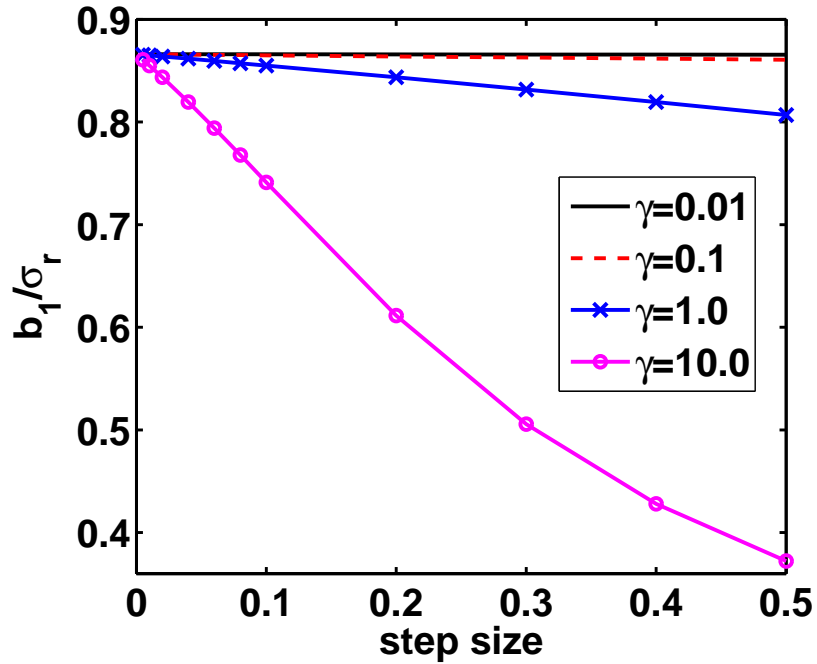}
\caption{(Color online) The correlation coefficient between the velocity and position vs.\ the time step size, for
$\gamma=0.01$, $0.1$, $1.0$ and $10.0$, respectively. } \label{fig_cov}
\end{figure}

We have provided detailed derivation of a class of Gear-like PC methods for performing BD simulation, and we tested them by
using the SDHO model. In particular, we have compared the performances of different BD methods, namely, the Euler-like,
Beeman-like (Verlet-like) and Gear-like methods, and discussed several similarities and differences between the BD and MD
simulations.

However, it may be sometimes computationally too demanding to implement a full BD simulation scheme, so we discuss briefly here
some possible derivatives of the BD method.

One possible simplification that might be made to the above BD simulation procedure is to neglect the correlation between the
velocity and position, i.e., to neglect their covariance. This may be done by simply setting $b_{1}=0$ in Eq.\ (\ref{equpdate}),
as many researchers actually do in their simulations. Therefore, we have also performed simulations without the covariance, and
the results turned out essentially the same as those taking full account of the covariance. Physically, it is not easy to see
why such a simplification should work at all. In fact, Fig.\ \ref{fig_cov}, showing the correlation coefficient
$b_{1}/\sigma_{r}$ versus the time step size for different $\gamma$, implies that the correlation between the velocity and
position is not negligible at all. Therefore, it seems that monitoring the accuracy, energy conservation, and statistical
properties of the ensembles of data from numerical results is not sufficient to fully evaluate performance of an algorithm.
Further simulations with many-particle systems are currently being carried out to test if this simplification would affect any
other physical quantities of such systems.

Yet another possible simplification to the above procedure is to apply a finite difference scheme for the deterministic force
directly to Langevin equation (\ref{eqlangevin}), and numerically solve the resultant formulae like in MD simulation, but with
the inclusion of corrections from the Langevin terms at the end of each time step. It can be proved that this simplification is
equivalent to using linear approximations to the coefficients $c_{n}$ in the Taylor-like series of Eq.\ (\ref{eqglmean}), for
example, by letting
\begin{eqnarray}
c_{0} &\approx& 1-\gamma t, \nonumber \\
c_{1} &\approx& 1-\frac{1}{2}\gamma t, \nonumber \\
c_{2} &\approx& \frac{1}{2}-\frac{1}{6}\gamma t, \nonumber \\
c_{3} &\approx& \frac{1}{6}-\frac{1}{24}\gamma t \nonumber \\
\cdots
\end{eqnarray}
Clearly, when $\gamma t\ll 1$, the above expressions would provide very good approximations to their real values given by Eq.\
(\ref{eqglmean}). However, even if $\gamma t=0.01$, the use of this approximation would produce an error of magnitude around
$10^{-4}$, which is still much larger than the accuracy that the Gear-like methods can achieve.

\subsection{Comparison with High-order Runge-Kutta-like methods}
We provide here a brief discussion of the strengths and weaknesses of our algorithm in comparison with some of the most recent
developments in the area of high-order algorithms for BD simulation, emphasizing that they typically involve Runge-Kutta methods
\cite{Hershkovitz1998,Drozdov1998,Forbert2000},

One should note that the algorithms based on decomposition of the exponential operator for time propagation provide superior
stability of simulation, especially in the limit of low damping, owing to their unique properties of conserving the phase-space
volume and being time reversible when simulating Hamiltonian dynamics problems. However, according to Bussi and Parrinello
\cite{Bussi2007}, while such algorithms are derived with the aim of producing accurate trajectories up to a given order, they
usually break down when a high damping is applied, which is the situation where time reversibility is irrelevant. Moreover, the
design of such algorithms is not focused on the correctness of the ensemble generated in simulation \cite{Bussi2007}, which may
become a serious shortcoming when simulating, e.g., equilibrium structure of a many-particle system interacting with a bath. On
the other hand, our method has demonstrated excellent performance in the long time limit, where equilibrium distribution
functions were reproduced within the statistical error, both for the present single-particle model system, and for the
many-particle applications in dusty plasmas \cite{Hou2006,LJH}.

An additional advantage of the methods based on the time propagator with Fokker-Planck operator lies in the fact that various
decomposition techniques make analysis quite transparent and allow for systematic strategies in developing higher-order schemes,
the highest so far being the fourth-order numerical integrators developed by Drozdov and Brey \cite{Drozdov1998}, and by Forbert
and Chin \cite{Forbert2000}. However, it is precisely the decomposition of the exponentiated Fokker-Planck operator that seems
to hinder further extensions to orders higher than fourth because of excessive growth in the complexity of such techniques with
increasing order. On the other hand, simplicity of our algorithm makes extensions to orders higher than fifth quite
straightforward, therefore offering an attractive pay-off for their lack of transparency.

As mentioned in the Introduction, the most serious limitation of algorithms involving high-order Runge-Kutta-like methods, such
as \cite{Hershkovitz1998,Drozdov1998,Forbert2000}, may arise when systems with large numbers of Brownian particles are
simulated, where the need for force evaluations including all inter-particle interactions becomes the most critical issue in any
simulation. For example, the method of Hershkovitz \cite{Hershkovitz1998}, when used in a 1D simulation, requires four
evaluations of force, four GRVs, and one evaluation of the force derivative per time step. Moreover, the method of Drozdov and
Brey \cite{Drozdov1998}, and the K4a method of Forbert and Chin \cite{Forbert2000} require three evaluations of force, one of
the force derivative, and four GRVs for the former method while the latter method requires even eight GRVs. The K4b and K4c
methods of Forbert and Chin \cite{Forbert2000} require eight force evaluations and three GRVs per step. Yet the accuracy of
these methods goes only up to the fourth order, whereas our method achieves easily the fifth-order accuracy with only one force
evaluation and three GRVs per time step, which presents its greatest advantage for many-particle simulations.

It should be mentioned, however, that the main weakness of our algorithm lies in the fact that it can handle only simple models
of particle interactions with a bath, characterized with constant friction (diffusion) coefficients and with fluctuations driven
by the Gaussian white noise. On the other hand, the methods based on the time propagator decomposition techniques
\cite{Drozdov1998,Forbert2000} are capable of solving rather general SDEs, e.g., multivariable Langevin equation with
configuration-dependent diffusion coefficients, or Kramers equation with colored noise.

\section{concluding remarks}
In summary, a Gear-Like Predictor-Corrector algorithm was proposed for BD simulation. It has been tested by using a 1D
stochastically-damped-harmonic-oscillator model in terms of its accuracy, energy conservation, and the long-time statistical
properties, and the results were compared with those obtained by using the Euler-like and Beeman-like (Verlet-like) methods. It
was found that the present method exhibits much better performance in all the above tests. At the same time, when compared to
the recent high-order Runge-Kutta-like methods \cite{Hershkovitz1998,Drozdov1998,Forbert2000}, our algorithm promises superior
efficiency for simulations in many-particle systems with constant friction (diffusion) coefficient and with Gaussian white
noise, as only one force evaluation is needed per time step.

We note that our Gear-like algorithm has already been used successfully for simulations of both equilibrium and nonequilibrium
phenomena in strongly-coupled dusty plasmas, involving up to $\sim 10^4$ interacting dust particles \cite{Hou2006,Hou2008}. More
simulations with many-particle Yukawa systems are being carried out to further test the performance of our method, and the
results will be reported elsewhere \cite{LJH}.

\begin{acknowledgments}
This work was supported by NSERC and PREA. L.J.H. is now at Christian-Albrechts Universit\"{a}t zu Kiel, and he gratefully
acknowledges supports from Alexander von Humboldt Foundation and from Prof. Alexander Piel.
\end{acknowledgments}

\appendix

\section{Stochastic damped harmonic oscillator}
The problem of a stochastic damped harmonic oscillator was first worked out in detail by Chandrasekhar in 1943
\cite{Chandrasekhar1943}. We provide here the list of his final results to facilitate the discussion in the main text.
\begin{eqnarray}
\text{mean}\{r(t)\} &=& r_{0}e^{-\gamma t/2} \left(\cosh{\frac{\omega_{1}t}{2}}+
\frac{\gamma}{\omega_{1}}\sinh{\frac{\omega_{1}t}{2}}\right) +
\frac{2v_{0}}{\omega_{1}}e^{-\gamma t/2}\sinh{\frac{\omega_{1}t}{2}}, \nonumber \\
\text{mean}\{v(t)\} &=& v_{0}e^{-\gamma t/2} \left(\cosh{\frac{\omega_{1}t}{2}} -
\frac{\gamma}{\omega_{1}}\sinh{\frac{\omega_{1}t}{2}}\right) -
\frac{2x_{0}\omega_{0}}{\omega_{1}}e^{-\gamma t/2}\sinh{\frac{\omega_{1}t}{2}} , \nonumber \\
\text{var}\{r(t)\} &=& \frac{k_{B}T}{m\omega_{0}^2} \left[1-e^{-\gamma t}
\left(1+2\frac{\gamma^2}{\omega_1^2}\sinh^2{\frac{\omega_{1}t}{2}}+
\frac{\gamma}{\omega_{1}}\sinh{\omega_{1}t}\right)\right], \nonumber \\
\text{var}\{v(t)\} &=& \frac{k_{B}T}{m} \left[1-e^{-\gamma t}
\left(1+2\frac{\gamma^2}{\omega_1^2}\sinh^2{\frac{\omega_{1}t}{2}}- \frac{\gamma}{\omega_{1}}\sinh{\omega_{1}t}\right)\right],
\nonumber \\
\text{cov}\{r(t)v(t)\} &=& \frac{4\gamma k_{B}T}{m\omega_1^2}e^{-\gamma t}\sinh^2{\frac{\omega_{1}t}{2}}, \label{eqsdho}
\end{eqnarray}
where $\omega_{1}=\sqrt{\gamma^2-4\omega_{0}^2}$. It should be noted that the above equations apply only when $\gamma\geq
2\omega_{0}$, i.e., in the over-damped case, while in the weakly-damped case, i.e., when $\gamma<2\omega_{0}$, one needs to
simply redefine $\omega_{1}$ as $\omega_{1}=\sqrt{4\omega_{0}^2-\gamma^2}$, and replace $cosh$ and $sinh$ with $cos$ and $sin$,
respectively.

One can easily see from the above expressions that, when $\gamma t \rightarrow \infty$, the solutions become:
\begin{eqnarray} r(t) & =&
\sqrt{\frac{k_{B}T}{m\omega_{0}^2}}N_{r}(0,1) \nonumber \\
v(t) &=& \sqrt{\frac{k_{B}T}{m}}N_{v}(0,1) , \label{eqsdhon}
\end{eqnarray}

\end{document}